\documentclass[aps,prb,showpacs,superscriptaddress,twocolumn]{revtex4}
\usepackage{amsmath}
\usepackage{amsfonts}
\usepackage{graphicx}
\usepackage{psfrag}
\usepackage{color}
\usepackage{bbold}

\begin{document}

\title{Oscillating spin-orbit interaction in two-dimensional superlattices: \\sharp transmission resonances and time-dependent spin polarized currents}

\author{Viktor Szaszk\'{o}-Bog\'{a}r}
\email{vszaszko@physx.u-szeged.hu}
\affiliation{Department of Theoretical Physics, University of Szeged, Tisza Lajos k\"{o}r%
\'{u}t 84, H-6720 Szeged, Hungary}
\affiliation{Departement Fysica, Universiteit Antwerpen, Groenenborgerlaan 171, B-2020
Antwerpen, Belgium}
\author{F. M. Peeters}
\affiliation{Departement Fysica, Universiteit Antwerpen, Groenenborgerlaan 171, B-2020
Antwerpen, Belgium}
\author{P\'{e}ter F\"{o}ldi}
\affiliation{Department of Theoretical Physics, University of Szeged, Tisza Lajos k\"{o}r%
\'{u}t 84, H-6720 Szeged, Hungary}

\begin{abstract}
We consider ballistic transport through a lateral, two-dimensional superlattice with experimentally realizable, sinusoidally oscillating Rashba-type spin-orbit interaction. The periodic structure of the rectangular lattice produces a spin-dependent miniband structure for static SOI. Using Floquet theory, transmission peaks are shown to appear in the mini-bandgaps as a consequence of the additional, time-dependent SOI. A detailed analysis shows that this effect is due to the generation of harmonics of the driving frequency, via which e.g., resonances that cannot be excited in the case of static SOI become available. Additionally, the transmitted current shows space and time-dependent partial spin-polarization, in other words, polarization waves propagate through the superlattice.
\end{abstract}

\maketitle

\section{Introduction}
Ballistic transport phenomena governed by time-dependent potentials are of fundamental interest mainly due to their close relation to important time-dependent quantum mechanical scattering effects. On the other hand, the possibility of controlling  the electron dynamics using time-dependent gate voltages may result in practical applications. Additionally, as it has been demonstrated experimentally, spin-dependent properties of semiconductor materials -- that are of exceptional importance e.g. in the context of spin-based quantum mechanical information processing\cite{ABD13,ZFS04} -- can also be controlled by gate voltages.\cite{N97,G00} These results motivated us to investigate how oscillating Rashba-type spin-orbit interaction (SOI)~\cite{R60} affects spin-dependent conductance in two-dimensional superlattices.
\bigskip

Spin-dependent transport phenomena in lateral superlattices~\cite{MM12,ZLQ12,LRZZ13} have been investigated experimentally, mainly in a two-dimensional network of quantum rings.\cite{KNAT02,BKSN06} Control of spin geometric phase in semiconductor quantum rings has also been demonstrated.\cite{NTKKN12,NFSRN13} Here, we focus on the ballistic regime and consider rectangular geometries, i.e., networks that consist of linear quantum wire segments as building blocks (see Fig.~\ref{latticefig}). The (quasi)periodic geometry of these devices results in a Rashba spin-orbit interaction controlled miniband structure,\cite{FSP10} with characteristic energies orders of magnitude below the usual band widths. This is a direct consequence of the difference between the lattice constant $a$ (see Fig.~\ref{latticefig}), the order of $100$ nm, and typical atomic separations. Since the position, width and even the existence of the non-conducting energy ranges (i.e., the mini-bandgaps) can be controlled experimentally via the strength of the SOI interaction, the conductance of the device is found to be tunable even at room temperature.\cite{FSP11}

In the current paper we demonstrate that the time dependence of the spin-orbit interaction gives rise to new physical phenomena, leading to observable transmission peaks in the mini-bandgaps. We consider the combination of oscillating and static SOI, where the latter one determines the miniband structure, while the oscillating part induce time-dependent effects. Note that transport related problems with oscillating SOI (but without the miniband structure) have been studied in Refs.~[\onlinecite{WC07,FBKP09}] for a ring, and in Ref.~[\onlinecite{RCM08}] for a ring-dot system. Application of Floquet's theory~\cite{F883} allows us to obtain nonperturbative results, high order harmonics of the SOI oscillation frequency appear in the transmission. Floquet scattering matrix theory is proven to be a useful mathematical tool for the description of periodically time dependent phenomena in diverse mesoscopic samples.\cite{LR99,LSS12,M14,DFB14} Specifically, several studies have discussed resonant phenomena of quantum dots and nanowires in the presence of time-dependent potential, see e.g.~Refs~[\onlinecite{NSzP012,SFP14,SS13,PSzSB14}].

Here, we will show that from a detailed analysis, we find the higher harmonics of the SOI oscillation frequency are responsible for the transmission peaks in the mini-bandgaps e.g., by allowing the excitation of resonances that are not coupled to the input/output leads for static SOI.

The present paper is organized as follows. In section \ref{mocdelsec}, we describe the model and methods that are used in the following.  Physical consequences of this model are analyzed in Sec.~III. We present and discuss our time averaged results in section IV. Time resolved spin and charge density oscillations are discussed in Sec.~V, while Sec.~VI contains the summary and conclusions.

\section{Model and methods}
\label{mocdelsec}
\begin{figure}[tbh]
\includegraphics[width=8cm]{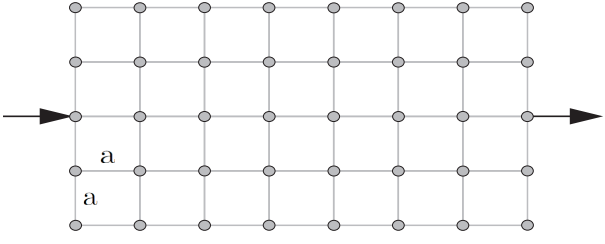}
\caption{Schematic view of the rectangular two-dimensional superlattice. Quantum wires represented by gray lines are subjected to oscillating Rashba-type SOI. Black arrows show the input and output leads in which no spin-orbit coupling is present.}
\label{latticefig}
\end{figure}

The building blocks of the superlattices shown in Fig.~\ref{latticefig} are linear, narrow (one dimensional) quantum wires. The corresponding Hamiltonian with Rashba-type SOI can be written \cite{FKB10,TELE12,MMK02,KPV11,ZZ13} as
\begin{equation}
H(\tau) = \hbar\Omega\left[ \left( -i\frac{\partial}{\partial s}+\frac {\omega(\tau)}{2\Omega} \mathbf{n}(\mathbf{\sigma}\times\mathbf{e}_z) \right) ^{2}-\frac{\omega(\tau)^{2}}{4\Omega^{2}}
\right],
\label{Ham}
\end{equation}
where the unit vector $\mathbf{n}=\cos \varphi \hat{\mathbf{e}}_x + \sin \varphi \hat{\mathbf{e}}_y$ points to the chosen positive direction along the wire, $s$ denotes the dimensionless position variable (measured in units of the lattice constant, $a$) and we introduced the characteristic kinetic energy $\hbar\Omega=\hbar^{2}/2m^{\ast}a^{2}.$  Note that analogous Hamiltonian has also been used previously for quantum rings.\cite{Sz11,PZ12,SSzEP12,Ma15,KFBP08b,KFBP08c,EE14} The strength of the SOI is given by $\omega(\tau)=\alpha(\tau)/a,$ where $\tau=\Omega t$ is the (dimensionless) time variable, and $\alpha$ denotes the Rashba parameter. The time dependence of $\omega(\tau)$ is assumed to be given by
\begin{equation}
\omega(\tau)=\omega_{0}+\omega_{1}\cos(\nu_{\alpha}\tau),
\label{alphat}
\end{equation}
where $\nu_{\alpha}$ is the frequency of the gate voltage oscillation (in units of $\Omega$).

\subsection{Floquet states and quasienergies}
\label{floquetsubsec}
Considering the solution of the time-dependent Schr\"{o}dinger equation
\begin{equation}
i\frac{\partial \Phi(s,\tau)}{\partial \tau} = \frac{1}{\hbar\Omega}{H}(\tau)\Phi(s,\tau),
\label{TDSE}
\end{equation}
it is seen that for an infinite, narrow quantum wire, any initial state can be expanded as a linear combination (a continuous one in case of the spatial variable) of spinor valued wave functions
\begin{equation}
\Phi_1(k,s)=e^{i k s} \left(\begin{array}{cc}
1 \\
0
\end{array} \right),
\
\Phi_2(k,s)=e^{i k s} \left(\begin{array}{cc}
0 \\
1
\end{array} \right),
\label{bases}
\end{equation}
which are expressed in the eigenbasis of $\sigma_z.$ For a given value of $k$ (measured in units of $1/a$), the action of the Hamiltonian on the states (\ref{bases}) becomes relatively simple, since the spatial derivatives have to  be replaced by a multiplication by $ik.$ This, together with the fact that
\begin{equation}
\left[H(\tau),H(\tau')\right]=0,
\label{commh}
\end{equation}
for any two time instants $\tau$ and $\tau',$ allows us to calculate the time evolution for an arbitrary initial state. Concretely, the evolution operator, for which
\begin{equation}
U(k,\tau) \left[ \Phi(k,s,\tau=0)\right] = \Phi(k,s,\tau)
\label{U}
\end{equation}
for any linear combination $\Phi(k,s,\tau=0)=\alpha\Phi_1(k,s)+\beta\Phi_2(k,s),$ can be calculated explicitly:
\begin{align}
U(k,\tau)=e^{-ik^{2}\tau}\times&\left[\mathbb{1}\cos\left(\frac{\omega_{0}}{\Omega}k\tau+\frac{\omega_{1}k}{\Omega\nu_{\alpha}}\sin(\nu_{\alpha}\tau)\right) \right. +  \nonumber \\
&\left. \sigma_\varphi\sin\left(\frac{\omega_{0}}{\Omega}k\tau+\frac{\omega_{1}k}{\Omega\nu_{\alpha}}\sin(\nu_{\alpha}\tau)\right) \right].
\label{Umat}
\end{align}
Here, $\mathbb{1}$ denotes the $2\times 2$ identity matrix, while $\sigma_\varphi=\cos \varphi \sigma_x + \sin \varphi \sigma_y$ is the Pauli matrix corresponding to the direction of the actual lead, with $\varphi$ representing the appropriate polar angle [for $\mathbf{n}=\hat{\mathbf{e}}_x \ (\hat{\mathbf{e}}_y)$, $\varphi=0\ (\pi/2)$].

The time evolution operator (\ref{Umat}) is diagonal in the basis of
\begin{equation}
\psi_\pm(k,s,\tau)=\frac{1}{\sqrt{2}}e^{i \left[k s-\epsilon_\pm(k,\tau)\right]} \left(\begin{array}{cc}
1 \\
\pm i e^{i \varphi}
\end{array} \right),
\label{tdsols}
\end{equation}
where
\begin{equation}
\epsilon_\pm(k,\tau)=(k^{2}\pm \frac{\omega_{0}}{\Omega}k)\tau\mp i\frac{\omega_{1}k}{\Omega\nu_{\alpha}}\sin(\nu_{\alpha}\tau).
\label{Floq}
\end{equation}
Note that the eigenspinors of the spin operator $\sigma_{\varphi}$ have the form
\begin{equation}
|\varphi_{\pm}\rangle = \frac{1}{\sqrt{2}}\left(\begin{array}{cc}
1 \\
\pm ie^{i\varphi}
\end{array} \right).
\end{equation}
This means that the time-dependent basis spinors can be written as
\begin{equation}
\psi_\pm(k,s,\tau)=U(k,\tau) \left[ \psi_\pm(k,s,0)\right]=e^{-i\epsilon_\pm(k,\tau)} \psi_\pm(k,s,0).
\end{equation}
The exponential factor above can be factorized:
\begin{equation}
e^{-i\epsilon_\pm(k,\tau)}=e^{-i\epsilon_\pm^0(k)\tau}e^{\mp i\frac{\omega_{1}k}{\Omega\nu_{\alpha}}\sin(\nu_{\alpha}\tau)},
\label{eigentd}
\end{equation}
where
\begin{equation}
\epsilon_\pm^0(k)=k^{2}\pm \frac{\omega_{0}}{\Omega}k,
\label{eps0}
\end{equation}
and the second term (that is periodic in time) can be expanded as:
\begin{equation}
e^{\mp i\frac{\omega_{1}k}{\Omega\nu_{\alpha}}\sin(\nu_{\alpha}\tau)}=\sum_{n=-\infty}^{+\infty}J_{n}\left(\frac{\omega_{1}k}{\Omega\nu_{\alpha}}\right)e^{\pm in\nu_{\alpha}\tau}.
\label{Jacobi}
\end{equation}
The Jacobi-Anger identity above (where Bessel functions of the first kind~\cite{AS65} appear), explicitly shows that the states
 $\psi_\pm(k,s,\tau)$ can be called the Floquet states of the problem corresponding to the quasienergies of $\epsilon_\pm^0(k).$

\subsection{Global solution of the transport problem}
\label{transportsubsec}
Having obtained the "time-dependent eigenspinors" (\ref{tdsols}) of the Hamiltonian (\ref{Ham}), we have the solution of the time-dependent Schr\"{o}dinger equation (\ref{TDSE}) for an arbitrary initial state in an infinite quantum wire. The transport problem of the superlattice, however, involves quantum wire segments, and the solution has to obey appropriate boundary conditions.

We assume that a monoenergetic plane wave input enters the network from the left hand side (see Fig.~\ref{latticefig}):
\begin{equation}
\psi_{in}(s,\tau)=\frac{1}{\sqrt{2}}e^{i \left[k_{in}(E_0) x-E_0 \tau)\right]} \left(\begin{array}{cc}
a \\
b
\end{array} \right),
\label{input}
\end{equation}
where, in dimensionless units,
\begin{equation}
k_{in}(E)=\sqrt{E}.
\label{kfree}
\end{equation}
The oscillating part of the SOI can induce high harmonics of frequency $\nu_\alpha$, leading to "sidebands" or Floquet channels~\cite{WC07} in the transmission at dimensionless energies
\begin{equation}
E_n=E_0+n\nu_\alpha,
\label{energies}
\end{equation}
with $n$ being integer. Note that although SOI oscillations are obviously not quantized, this expression resembles the scenario when a quantum system absorbs/emits energy quanta proportional to $\nu_\alpha$ from/into a quantized field. The appearance of the frequencies (\ref{energies}) in the time evolution of the quantum state of the system means that e.g., the reflected spinor valued wave function can be written as
\begin{equation}
\psi_{refl}(s,\tau)=
\sum_n e^{i \left[-k_{in}(E_n) s-E_n \tau)\right]} \left(\begin{array}{cc}
r_{1n} \\
r_{2n}
\end{array} \right),
\label{reflection}
\end{equation}
where the coefficients $r_{1n}$ and $r_{2n}$ will be determined by the boundary conditions. Similarly, for the transmission (in the output arm)
\begin{equation}
\psi_{trans}(s,\tau)=
\sum_n e^{i \left[k_{in}(E_n) s-E_n \tau)\right]} \left(\begin{array}{cc}
t_{1n} \\
t_{2n}
\end{array} \right).
\label{transmission}
\end{equation}
As we shall see, the complete transport problem can be solved by imposing appropriate boundary conditions at the junctions -- in frequency domain, that is, for each frequency component (\ref{energies}) separately. By investigating Eqs.~(\ref{tdsols})-(\ref{Jacobi}), one can see that the time evolution of the solutions (\ref{tdsols}) involves a given frequency $E_n,$ whenever
\begin{equation}
\epsilon_\pm^0(k)=E_m,
\label{epseqEn}
\end{equation}
where $m$ and $n$ can be either equal or different. [In fact, once a term $\exp(-iE_n\tau)$ appears in the time evolution of a state given by Eq.~(\ref{tdsols}), all other frequencies $E_m=E_n+(m-n)\nu_\alpha$ play a role -- although it is possible that their weight in the Fourier expansion is negligible.] By solving Eqs.~(\ref{epseqEn}) for the wave number $k$, we obtain that
\begin{eqnarray}
k^{1,2}_{SOI}(E_m)=-\frac{\omega_1}{2\Omega} \pm \sqrt{\frac{\omega_1^{2}}{4\Omega^{2}}+E_m}, \nonumber \\
k^{3,4}_{SOI}(E_m)=\frac{\omega_1}{2\Omega} \pm \sqrt{\frac{\omega_1^{2}}{4\Omega^{2}}+E_m},
\label{ksoi}
\end{eqnarray}
where the first two solutions correspond to the upper, while $k^{3,4}_{SOI}$ to the lower sign in Eq.~(\ref{epseqEn}), and the subscript reminds us that these relations are valid in domains with oscillating SOI interaction. [Note the difference between Eqs.~(\ref{kfree}) and (\ref{ksoi}).]
Combining Eqs.~(\ref{tdsols}), (\ref{epseqEn}) and (\ref{ksoi}), we see that a general solution of the time-dependent Schr\"{o}dinger equation (\ref{TDSE}) that involves the frequency components (\ref{energies}) can be written in the following form:
\begin{equation}
\psi_{SOI}(s,\tau)=
\sum_{m=-\infty}^{\infty} \sum_{i=1}^{4}a^{i}\psi_{i}\left(k^{i}_{SOI}(E_m),s,\tau\right),
\label{inSOI}
\end{equation}
where the coefficients $a_i$ will have to be determined using the boundary conditions, and
\begin{align}
 \psi_{m,i}(k,s,\tau) = \left\{
 \begin{array}{l l}
    \psi_+\left(k^{i}_{SOI}(E_m),s,\tau\right) & \quad \text{for $i=1,2$}\\
    \newline\\
    \psi_-\left(k^{i}_{SOI}(E_m),s,\tau\right) & \quad \text{for $i=3,4$ .}
  \end{array}\right.
 \label{insoihelp}
\end{align}
Using the Jacobi-Anger expansion (\ref{Jacobi}), we obtain e.g.:
\begin{align}
\nonumber
\psi_{m,1}\left(k,s,\tau \right)=&\psi_{m,1}\left(k,s,0 \right)\\
\times& \sum_{l=-\infty}^{\infty}e^{-i\tau E_{m-l}}J_{l}\left(\frac{\omega_{1}k}{\Omega\nu_{\alpha}}\right).
\label{expansion}
\end{align}

\bigskip

At this stage, we obtained solutions to the time-dependent Schr\"{o}dinger equation in all spatial domains: $\psi_{in}(x,\tau)+\psi_{refl}(x,\tau)$ in the input arm, $\psi_{trans}(x,\tau)$ in the output arm, and $\psi_{SOI}(s,\tau)$ (with appropriate orientation, i.e., value of $\varphi$) inside the network. These spinor valued wave functions contain coefficients that are still to be determined. Using the coordinate system shown in Fig.~\ref{latticefig}, we require $\mathrm{Re}(k_{in})>0,$ $\mathrm{Im}(k_{in})<0$ in the input arm, and $\mathrm{Re}(k_{in})>0,$ $\mathrm{Im}(k_{in})>0$ for the transmitted solution, in order to ensure left propagating reflected waves, right propagating transmitted waves and evanescent solutions that decay as a function of the distance from the central region. Since the domains on which the functions $\psi_{SOI}(s,\tau)$ are defined are finite, and propagation in both directions is possible, there are no restriction for $k^i_{SOI}.$ At the junctions (input, output and internal ones) we apply Griffith's boundary conditions~\cite{G53} for each frequency component separately. In this way the spinor valued wave function is continuous at any time instant at all the junctions, and the net spin current density that leaves/enters any given junction disappear always. The resulting infinite system of linear equations (for more details, see the Appendix) can be truncated. Practically, for the results presented in this paper, it turns out that considering approximately 50 values of $E_n$ ($n=-25, \ldots 25$) is sufficient to achieve accurate results. This can be checked reliably via calculating the time averaged transmission and reflection probabilities (see the next section) that has to add up to unity. When this sum is not close enough to 1 (within $10^{-5}$ relative error), we increase the number of frequency components that are taken into account.

\begin{figure}[tbh]
\centering
\includegraphics[width=8cm]{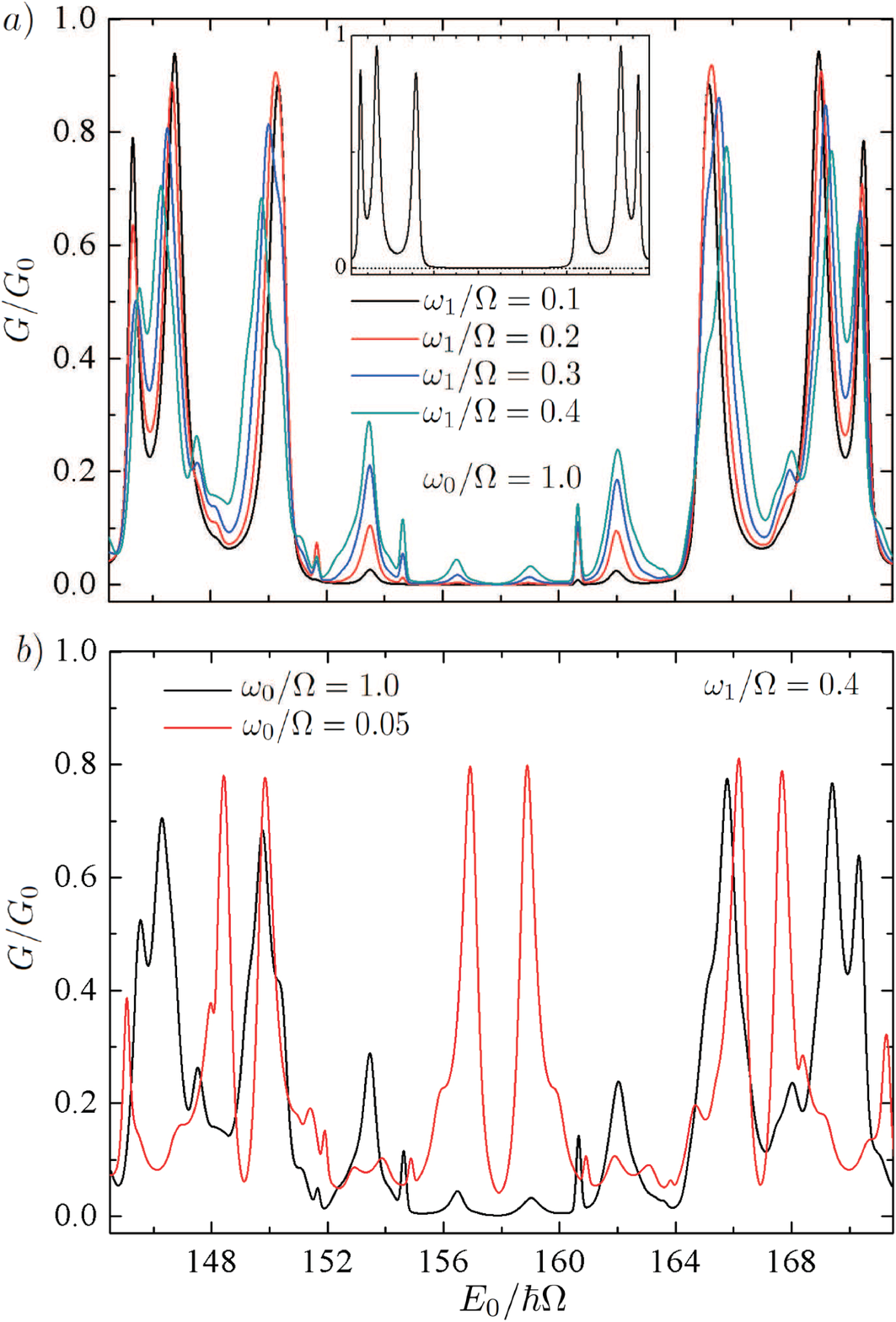}
\caption{(Color online) Conductance (in units of $G_0=\frac{2e^{2}}{h}$) of a $7\times 7$ array in the presence of oscillating and stationary Rashba-type spin-orbit interaction [see Eq.~(\ref{alphat})]. Panel a) focuses on an energy range that corresponds to a mini-bandgap without the oscillating part of the SOI (see the inset). In this panel, $G$ is plotted for different oscillating SOI amplitudes $\omega_1,$ with $\omega_0$ being fixed. According to panel b), when $\omega_{0}/\Omega$ is considerably lower than $\omega_{1}/\Omega$, the mini-bandgap disappears. Parameters are: $\omega_{0}/\Omega=1.0$, $\nu_{\alpha}=3.0$ (panel $a)$) and $\omega_{1}/\Omega=0.4$, $\nu_{\alpha}=3.0$ (panel $b)$).}
\label{peak}
\end{figure}

\section{Observables}
\label{observablesubsec}
The results to be presented in the current section are related to physical quantities that can be calculated using the solution
 \begin{equation}
\psi(s,\tau)=
\left(\begin{array}{cc}
u_1(s,\tau) \\
u_2(s,\tau)
\end{array} \right).
\label{generalsol}
\end{equation}
[That stands for $\psi_{in}(x,\tau)+\psi_{refl}(x,\tau),$ $\psi_{trans}(x,\tau)$ or $\psi_{SOI}(s,\tau),$ depending on the location].
The position dependent (unnormalized) electron density is given by
 \begin{equation}
n(s,\tau)=\psi^{\dagger}(s,\tau)\psi(s,\tau)=|u_1(s,\tau)|^2+|u_2(s,\tau)|^2,
\end{equation}
while the density for the spin-up and spin-down electrons (in the $z$ direction) reads
\begin{equation}
n_{\uparrow}(s,\tau)=|u_1(s,\tau)|^2, \ \ n_{\downarrow}(s,\tau)=|u_2(s,\tau)|^2.
\end{equation}
Focusing on the spin degree of freedom, one can construct the quantum mechanical spin density operator
\begin{equation}
\rho(s,\tau)=\frac{1}{n(s,\tau)}
\left(\begin{array}{cc}
|u_1(s,\tau)|^2 & u_1(s,\tau) u_2^*(s,\tau) \\
u_1^*(s,\tau) u_2(s,\tau) & |u_2(s,\tau)|^2
\end{array} \right),
\label{generalsol2}
\end{equation}
which is defined only for nonzero electron density. Note that $\mathrm{Tr}[\rho]=1$ (by construction), and for spin polarized states $\mathrm{Tr}[\rho^2]=1$ also holds. However, when we would like to perform calculations for completely unpolarized input, the easiest way is to consider two different $\psi_{in}$ states separately, with their spinor parts being antiparallel, and finally add the results incoherently, with equal statistical weight (i.e., 1/2) being associated to each states. Formally, this is equivalent to an input spin density operator that is 1/2 times the $2\times 2$ identity matrix [see Eq.~(\ref{unpolinput})]. In this case $\mathrm{Tr}[\rho^2_{in}]=1/2,$ suggesting that the quantity $\mathrm{Tr}[\rho^2(s,\tau)]$ can be an appropriate local measure of spinpolarization (As it is often used in different contexts, see e.g. Ref.~[\onlinecite{FCB01a}]). However, in our case it is more intuitive to express the degree of spin polarization as the length of the vector
\begin{equation}
\hat{\mathbf{S}} (s,\tau)=
\left(\begin{array}{cc}
\mathrm{Tr}[\rho(s,\tau)\sigma_x] \\
\mathrm{Tr}[\rho(s,\tau)\sigma_y] \\
\mathrm{Tr}[\rho(s,\tau)\sigma_z]
\end{array} \right)
\label{orientation}
\end{equation}
that describes the spin orientation (as the components are the expectation values of the Pauli matrices). As it can be shown easily, $\hat{\mathbf{S}}$ is a unit vector for spin polarized states, while its length is zero when $\rho$ is proportional to unity. The "purity"
\begin{equation}
p=\sqrt{\hat{\mathbf{S}}\hat{\mathbf{S}}}
\end{equation}
will be used to measure the degree of spin polarization. (Note that $p\in[0,1].$)

As an important quantity that does not depend on time, we calculate the time averaged transmission probability,
\begin{equation}
\mathcal{T}=\frac{1}{J_{in}}\int_0^T J_{out}(x=0,t) dt,
\end{equation}
where the usual quantum mechanical probability current densities appear, and $T=2\pi/\nu_\alpha.$ Explicit calculation using Eqs.~(\ref{input}) and (\ref{transmission}) shows that the time averaged conductance, which is proportional to $\mathcal{T},$ can be written in units of $2e^2/h$ as:
\begin{equation}
G=\frac{1}{(|a|^2+|b|^2)k_{in}(E_0)}\sum_n (|t_{1n}|^2+|t_{2n}|^2)k_{in}(E_n).
\label{avcond}
\end{equation}

\begin{figure}[tbh]
\centering
\includegraphics[width=8cm]{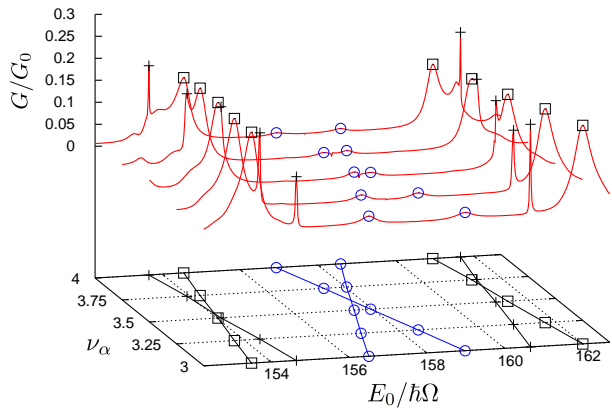}
\caption{(Color online) Conductance (in units of $G_0=\frac{2e^{2}}{h}$) of a $7\times 7$ array as a function of the input energy $E_0$ and the frequency $\nu_\alpha$ of the oscillating part of the SOI. Additional parameters: $\omega_0/\Omega=1.0,$ $\omega_0/\Omega=0.3.$}
\label{peakmove}
\end{figure}

\section{Time averaged conductance properties in the mini-bandgaps}
\label{averagedsubsec}
Fig.~\ref{peak} shows the time averaged conductance [see Eq.~(\ref{avcond})] as a function of the input energy $E_0$ for different parameters. The role of the oscillating part of the SOI is clearly seen by comparing the inset and the main plot in panel (a): in the energy range corresponding to a mini-bandgap of the system with $\omega_1=0,$ $G$ is practically zero for the static system, but the oscillating part of the SOI induces several conductance peaks. In the following we will analyse these peaks.

Let us recall\cite{FSP10} that there are no bandgaps without static SOI ($\omega_0=0$), and -- as a rule of thumb -- the widths of these energy ranges with zero conductance increase as a function of $|\omega_0|.$ As panel b) of Fig.~\ref{peak} shows, sufficiently strong oscillating SOI can smear out the miniband structure, even in cases when mini-bandgaps would still exist for $\omega_1=0.$ This is related to the fact that the larger the magnitude of $\omega_1$ is, the more pronounced the peaks in Fig.~\ref{peak} (a) are: broader and higher peaks in the mini-bandgaps lead to the disappearance of the bandgap itself.  As we shall see later, the peaks seen in Fig.~\ref{peak} (a) are related to the conductance via the Floquet channels corresponding to the harmonics (\ref{energies}), i.e., their appearance is a nonlinear effect. Therefore they are expected to play a more important role as the amplitude of the SOI oscillations -- that generate them -- increases.

\begin{figure}[tbh]
\centering
\includegraphics[width=8cm]{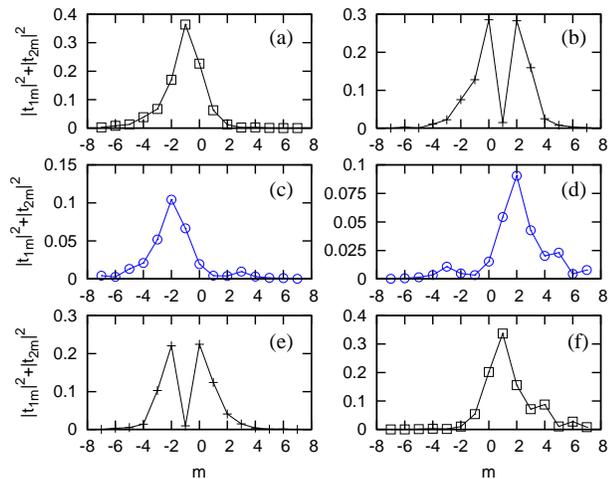}
\caption{(Color online) The weights $|t_{1m}|^2+|t_{1m}|^2$ of the frequency components $E_m$ [see Eqs.~(\ref{energies}) and (\ref{transmission})] in the transmitted spinor valued wave function as a function of the harmonic order (the integer $m$). Figures (a),(b) \ldots (f) correspond to the 6 peaks that can be identified in the front curve ($\nu_\alpha=3$) of Fig.~\ref{peakmove} -- from left (lowest energy) to the right (highest energy), respectively. For the clear identification of the peaks, the same symbols were used as in Fig.~\ref{peakmove}. Additional parameters are: $\omega_0/\Omega=1.0,$ $\omega_0/\Omega=0.3.$ }
\label{peakfreq}
\end{figure}

\bigskip
The position and physical origin of the conductance peaks in the mini-bandgaps needs a more detailed analysis. To this end it is instructive to see the dependence of the position of the peaks on the frequency $\nu_{\alpha}$ of the oscillating SOI. Fig.~\ref{peakmove} shows $G(E_0)$ for different values of $\nu_\alpha.$ Intuitively, based on their widths, heights, and shapes, we can identify three kinds of peaks in the mini-bandgap shown in Fig.~\ref{peakmove}. Local maxima that are similar in this sense are plotted using the same symbols and colors in this figure. As we shall see later, visual similarity corresponds to similar physical interpretation as well. First, focusing on the  projections on the bottom plane, we can see that the position of the local conductance maxima ($M$) changes linearly with $\nu_{\alpha}.$ More concretely, $M(\nu_{\alpha})=\mathrm{const}+n\nu_{\alpha},$ where $n$ has integer values that differ only in sign for the peaks that are plotted using the same symbols in Fig.~\ref{peakmove}. Although the "driving field" (Rashba-type SOI) is completely classical (the oscillations are not assumed to be quantized), this linearity, together with Fig.~\ref{peakfreq}, resembles the process of emission/absorption of oscillation quanta by the quantum system -- i.e., the electron. Consequently, Figs.~\ref{peakfreq} (a), (c), (d) and (f) -- that shows the weight of the frequency components (\ref{energies}) in the output give by Eq.~(\ref{transmission}) -- can be interpreted relatively easily. For panel (a) [(c)] the maximum of the transmission is shifted by (twice) $\nu_\alpha$ below $E_0.$ That is, the emergence of harmonics (sidebands around the input energy $E_0$) allows "mapping" of energy ranges with nonzero conductance into the mini-bandgap. Moreover, the position of peak (a) [(c)] has an energy (note that we are using dimensionless units) distance of $\nu_\alpha$ $(2\nu_\alpha)$ from the lower band edge around $E_0/\hbar\Omega=150$ (see Fig.~\ref{peakmove}). Similarly, peaks (d) and (f) "map" the nonzero conductance (that can be seen in Fig.~\ref{peak} (a) above $E_0/\hbar\Omega=165$) by the "absorption" of one [(f)] or two [(d)] "oscillation quanta" $\nu_\alpha.$

The narrow peaks whose energy distributions are denoted by (b) and (e) in Fig.~\ref{peakfreq}, are unrelated to the band edges, and their dependence on the driving frequency is different from that of peaks (a), (c), (d) and (f). According to Fig.~\ref{peakmove}, the energy difference between these peaks is $2 \nu_\alpha,$ to a very good approximation (within 2\% relative error). As we have checked by singular value decomposition (SVD) of the matrix that describes the boundary conditions (fitting at the junctions) for constant SOI ($\omega_1=0$), the energy value in the middle of these two peaks corresponds to a strong, multiply degenerate singular value. In other words, there are solutions that can be added to the scattering problem, that is, the global spinor valued wave function is not uniquely determined. However, these singular solutions have the property, that the corresponding electron densities are zero at the input junction. (E.g., for the parameters corresponding to Fig.~\ref{peakfreq}, the singularity appears at $E_0/\hbar\Omega=157.66,$ and there are standing probability waves around the input junction -- with a node being at this point -- so that the characteristic wavelength is $a/2$.) This means that these solutions are "closed", have no coupling to the input lead. In other words, singular solutions cannot be excited directly, i.e., not at the energy value where the matrix describing the boundary conditions is indeed singular without the oscillating part of the SOI. That is why the weight of the corresponding frequency component is practically zero in Figs.~\ref{peakfreq} (b) and (e). However, when nonlinear effects induced by the oscillating part of the SOI give rise to high harmonics, the corresponding wavelengths do not all result in destructive interference at the input junction: the conductance becomes nonzero.

Thus, all the peaks that appear in the mini-bandgap are related to the emergence of high harmonics of the frequency of the driving SOI oscillations, but the detailed physical mechanisms are different for the broad and narrow local conductance maxima. In the first case the edges of the mini conduction bands are "mapped" into the mini-bandgap, while a strong, narrow resonance is being excited in the latter case.

\section{Charge density oscillations and spin polarization}
\label{polarizationsubsec}
\begin{figure}[tbh]
\centering
\includegraphics[width=8cm]{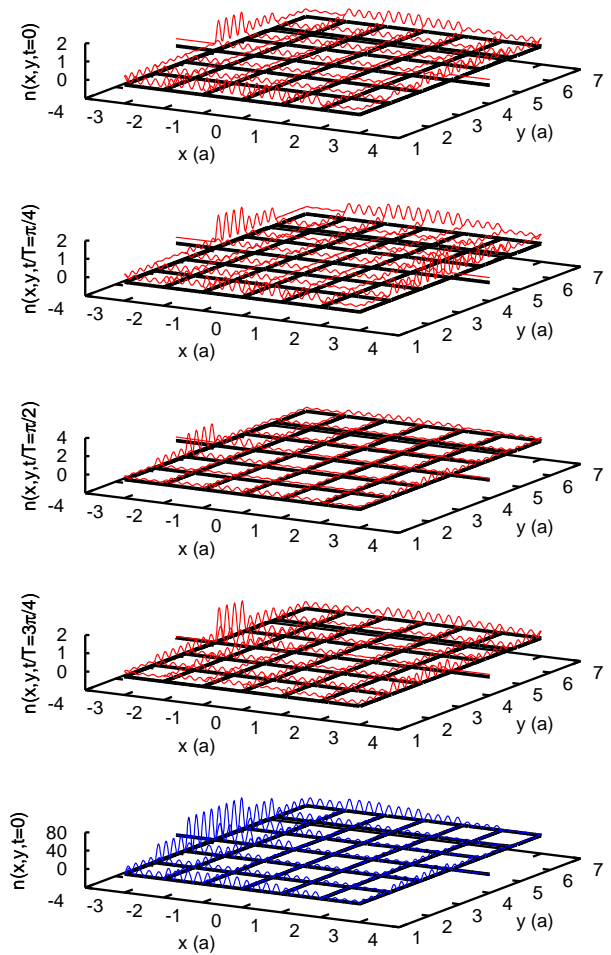}
\caption{(Color online) Snapshots of the electron density along a $7\times 7$ array for time instants indicated in the labels of the vertical axes. Note that $n$ is not normalized. [As a reference: the value of $n=1$ corresponds to the input plane wave (in this case with unpolarized spin state).] Panels (a)-(e) correspond to the first, broad peak seen in Fig.~\ref{peakmove} [panel (a) of Fig.~\ref{peakfreq}], while panel (e) correspond to the parameters of the second, narrow peak in Fig.~\ref{peakmove} [panel (b) of Fig.~\ref{peakfreq}].}
\label{tddens}
\end{figure}

Time-resolved details of the transmission can be visualized by plotting snapshots of the electron density along the network for various time instants. Here we consider completely unpolarized input, i.e., a plane wave with completely random spin polarization:
\begin{eqnarray}
\rho_{in}(x,\tau)&=&\frac{1}{2}e^{i \left[k_{in}(E_0) x-E_0 \tau)\right]}
\left [\left(\begin{array}{cc}
1 & 0 \\
0 & 0
\end{array} \right)+ \left(\begin{array}{cc}
0 & 0 \\
0 & 1
\end{array} \right)\right]\nonumber \\
&=&\frac{1}{2}e^{i \left[k_{in}(E_0) x-E_0 \tau)\right]}
\left(\begin{array}{cc}
1 & 0 \\
0 & 1
\end{array} \right)
.
\label{unpolinput}
\end{eqnarray}
As we can see in Fig.~\ref{tddens}, the density $n(x,y,\tau)$ has several maxima around the input junction, the location of which oscillates periodically during a cycle of duration $\mathcal{T}=2\pi/\nu_\alpha.$ This figure reveals an additional difference between the broad and the narrow peaks that appear in the mini-bandgaps: in the latter case [e.g., panels (b) and (e) in Fig.~\ref{peakfreq}], the excitation of an internal resonance of the network results in considerably higher particle densities.

Considering the output, it is instructive to point out that SOI oscillations can lead to temporal spinpolarization. This effect, in simpler geometries without miniband structure, has already been demonstrated,\cite{SFP14} thus our current findings -- besides providing a more detailed physical interpretation of the effect -- indicate that this is a general consequence of the time-dependent SOI, being essentially independent from the geometry of the system. Additionally, let us emphasize, that temporal spin-polarization is closely related to the oscillation of the SOI, no polarization appears for the case of static SOI. Without the time dependence of the spin-related properties of the device, strong, symmetry-based considerations\cite{KK05} related to the equilibrium spin currents rule out spin polarization effects.

\begin{figure}[tbh]
\centering
\includegraphics[width=8cm]{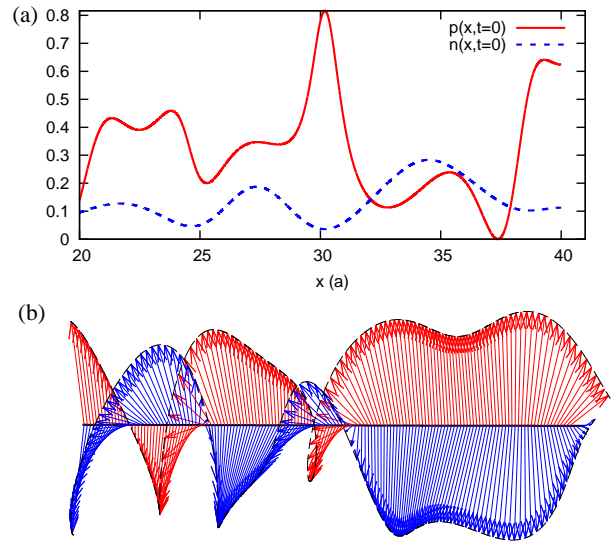}
\caption{(Color online) Panel a): The degree of spin polarization $p$ and the particle density $n$ in the output arm of a $7\times 7$ array at $\tau=0.$ The spin orientation corresponding to "spin-up" and "spin-down" (in the $z$ direction) input spinors are shown in panel b). Parameters are: $\omega_0/\Omega=1.0,$ $\omega_0/\Omega=0.3, $ $E_0/\hbar\Omega=153.5.$}
\label{purity}
\end{figure}

Figure \ref{purity} demonstrates that in the output arm, the degree of spin polarization characterized by $p$ can be close to unity such that the electron density is still nonzero. Let us note that spin polarization and density fluctuations appearing in this figure propagate away from the network in a wave-like manner. The arrows in Fig.~\ref{purity} (b) represent the spin orientation (\ref{orientation}) separately for the two, opposite input spin direction [see the first line of Eq.~(\ref{unpolinput})] the incoherent sum of which constitutes the input density matrix (\ref{unpolinput}). More precisely, the arrows visualize the spin direction in a local coordinate system, they point from $(x, 0, 0)$ to $(x+S_x, S_y, S_z).$ By investigating both panels of this figure, one can see the physical origin of the polarization effect: the spin directions corresponding to the two different input spinors rotate in a different way, they are not always antiparallel (which is the case for static SOI). In fact, there are spacetime points when these directions are almost the same, resulting in a remarkable partial polarization, $p.$ This emphasises the role of the oscillating part of the SOI in the spin polarization effect shown in Fig.~\ref{purity}.

\section{Conclusions}
We developed a model for the description of time and spin-dependent transport phenomena in rectangular, lateral superlattices. Motivated by recent experimental possibilities, the combined effect of static and oscillating Rashba-type spin-orbit interaction (SOI) were considered. The static part of the coupling induces an experimentally controllable miniband structure, while the oscillating part gives rise to transmission peaks in the mini-bandgaps. This effect is general for networks that contain more than $5\times5$ junctions. In order to see a clear miniband structure with relatively low computational costs -- without loss of generality -- we have chosen the size of $7\times 7$ to demonstrate our results.  We identified the physical mechanisms responsible for the appearance of conductance peaks in the mini-bandgaps, and have shown that the heights and positions of these peaks can be controlled by the amplitude and frequency of the SOI oscillations. These observations may lead to e.g.~narrow band, controllable energy filters.

\section*{Acknowledgments}
\vspace{-0.5cm}
This work was partially supported
by the European Union and the European Social Fund through projects
TAMOP-4.2.2.C-11/1/KONV-2012-0010) and TAMOP-4.2.2.A-11/1/KONV-2012-0060),
and by the Hungarian Scientific Research Fund (OTKA) under Contracts No.~T81364 and 116688.

\section*{Appendix}
As an example, let us consider the output junction (where the output lead is connected to the network). Using $s_0$ to denote the location of this point, Griffith' boundary conditions~\cite{G53} require the solution to be continuous at $s_0.$ That is, all the neighboring spinor valued wave functions evaluated at this point should be equal at any time instant. As an example, considering a quantum wire segment that joins the output junction and the output wire itself, we can write:
\begin{equation}
\psi_{SOI}(s=s_{0},\tau)=\psi_{trans}(s=s_{0},\tau),\nonumber
\end{equation}
\begin{align}
\sum_{m}
\Big\{&\big[a^{1}\psi_{+}(k^{1}_{SOI}(E_{m}),s_{0},\tau))  \nonumber \\
+&a^{2}\psi_{+}(k^{2}_{SOI}(E_{m}),s_{0},\tau))\big]\nonumber \\
+&\big[a^{3}\psi_{-}(k^{3}_{SOI}(E_{m}),s_{0},\tau))\nonumber \\
+&a^{4}\psi_{-}(k^{4}_{SOI}(E_{m}),s_{0},\tau))\big]\Big\}\nonumber\\
=\sum_{n} &e^{i \left[k_{in}(E_n) s_{0}-E_n \tau)\right]}
\left(\begin{array}{cc}
t_{1n} \\
t_{2n}
\end{array}\right),\nonumber
\end{align}
where the probability amplitudes $a^{i}$, the wavenumber $k^{i}_{SOI}(E_{m})$ and the states $\psi_{\pm}(k^{1}_{SOI}(E_{m}),s_{0},\tau))$ are introduced in Eqs.~(\ref{tdsols}), (\ref{ksoi}), (\ref{inSOI}) and (\ref{insoihelp}).

The periodicity (in time) of the problem offers a relatively simple way to imply the condition above, since (via Fourier series expansion) it is possible to work in the frequency domain. E.g., for the frequency component $E_m$ we have:
\begin{align}
\sum\limits_{l} \Big\{\left[a_{ml}^{1}e^{-ik^{1}_{SOI}(E_{m-l})s_{0}}+a_{ml}^{2}e^{-ik^{2}_{SOI}(E_{m-l})s_{0}}\right]|\varphi_{+}\rangle\qquad \nonumber \\
+ \left[a_{ml}^{3}e^{ik^{3}_{SOI}(E_{m+l})s_{0}}+a_{ml}^{4}e^{ik^{4}_{SOI}(E_{m+l})s_{0}}\right]|\varphi_{-}\rangle\Big\} \qquad\nonumber\\
\times J_{l}\left(\frac{\omega_{1}k}{\Omega\nu_{\alpha}}\right) = e^{ik_{in}(E_{m})s_{0}} \left(\begin{array}{cc}
t_{1m}(E_{m}) \\
t_{2m}(E_{m})
\end{array}\right).\label{cont}
\end{align}
The second part of the boundary conditions is related to the quantum mechanical probability current density, which, in the presence of the SOI, reads:
\begin{equation}
J(s,\tau)=2\mathrm{Re}\left\langle-i\frac{\partial}{\partial s}+\frac{\omega(\tau)}{2\Omega}\sigma_{\varphi}\right\rangle_{\Psi(s,\tau)},
\end{equation}
where $\Psi(s,\tau)$ denotes a solution to the time-dependent Schr\"{o}dinger equation (\ref{TDSE}). (A derivation that leads to an analogous expression for a ring, can be found in Ref.~[\onlinecite{MPV04}].) As one can check, having continuity imposed [Eqs.~(\ref{cont})], the condition that the net current density that flows in a junction (or, depending on the sign, out of it) should be zero at any time instant,\cite{G53} turns into a set of linear equations involving spatial derivatives. With an appropriate truncation of the infinite system of equations describing the boundary conditions, a global solution of the scattering problem can be achieved.


\begin{thebibliography}{43}
\expandafter\ifx\csname natexlab\endcsname\relax\def\natexlab#1{#1}\fi
\expandafter\ifx\csname bibnamefont\endcsname\relax
  \def\bibnamefont#1{#1}\fi
\expandafter\ifx\csname bibfnamefont\endcsname\relax
  \def\bibfnamefont#1{#1}\fi
\expandafter\ifx\csname citenamefont\endcsname\relax
  \def\citenamefont#1{#1}\fi
\expandafter\ifx\csname url\endcsname\relax
  \def\url#1{\texttt{#1}}\fi
\expandafter\ifx\csname urlprefix\endcsname\relax\def\urlprefix{URL }\fi
\providecommand{\bibinfo}[2]{#2}
\providecommand{\eprint}[2][]{\url{#2}}

\bibitem[{\citenamefont{Awschalom et~al.}(2013)\citenamefont{Awschalom,
  Bassett, Dzurak, Hu, and Petta}}]{ABD13}
\bibinfo{author}{\bibfnamefont{D.~D.} \bibnamefont{Awschalom}},
  \bibinfo{author}{\bibfnamefont{L.~C.} \bibnamefont{Bassett}},
  \bibinfo{author}{\bibfnamefont{A.~S.} \bibnamefont{Dzurak}},
  \bibinfo{author}{\bibfnamefont{E.~L.} \bibnamefont{Hu}}, \bibnamefont{and}
  \bibinfo{author}{\bibfnamefont{J.~R.} \bibnamefont{Petta}},
  \bibinfo{journal}{Science} \textbf{\bibinfo{volume}{339}},
  \bibinfo{pages}{1174} (\bibinfo{year}{2013}).

\bibitem[{\citenamefont{\v{Z}uti\'{c} et~al.}(2004)\citenamefont{\v{Z}uti\'{c},
  Fabian, and Sarma}}]{ZFS04}
\bibinfo{author}{\bibfnamefont{I.}~\bibnamefont{\v{Z}uti\'{c}}},
  \bibinfo{author}{\bibfnamefont{J.}~\bibnamefont{Fabian}}, \bibnamefont{and}
  \bibinfo{author}{\bibfnamefont{S.~D.} \bibnamefont{Sarma}},
  \bibinfo{journal}{Rev. Mod. Phys.} \textbf{\bibinfo{volume}{76}},
  \bibinfo{pages}{323} (\bibinfo{year}{2004}).

\bibitem[{\citenamefont{Nitta et~al.}(1997)\citenamefont{Nitta, Akazaki,
  Takayanagi, and Enoki}}]{N97}
\bibinfo{author}{\bibfnamefont{J.}~\bibnamefont{Nitta}},
  \bibinfo{author}{\bibfnamefont{T.}~\bibnamefont{Akazaki}},
  \bibinfo{author}{\bibfnamefont{H.}~\bibnamefont{Takayanagi}},
  \bibnamefont{and} \bibinfo{author}{\bibfnamefont{T.}~\bibnamefont{Enoki}},
  \bibinfo{journal}{Phys. Rev. Lett.} \textbf{\bibinfo{volume}{78}},
  \bibinfo{pages}{1335} (\bibinfo{year}{1997}).

\bibitem[{\citenamefont{Grundler}(2000)}]{G00}
\bibinfo{author}{\bibfnamefont{D.}~\bibnamefont{Grundler}},
  \bibinfo{journal}{Phys. Rev. Lett.} \textbf{\bibinfo{volume}{84}},
  \bibinfo{pages}{6074} (\bibinfo{year}{2000}).

\bibitem[{\citenamefont{Rashba}(1960)}]{R60}
\bibinfo{author}{\bibfnamefont{E.~I.} \bibnamefont{Rashba}},
  \bibinfo{journal}{Sov.~Phys.~Solid State} \textbf{\bibinfo{volume}{2}},
  \bibinfo{pages}{1109} (\bibinfo{year}{1960}).

\bibitem[{\citenamefont{Marinescu and Manolescu}(2012)}]{MM12}
\bibinfo{author}{\bibfnamefont{D.~C.} \bibnamefont{Marinescu}}
  \bibnamefont{and}
  \bibinfo{author}{\bibfnamefont{A.}~\bibnamefont{Manolescu}},
  \bibinfo{journal}{Phys. Rev. B} \textbf{\bibinfo{volume}{85}},
  \bibinfo{pages}{165302} (\bibinfo{year}{2012}).

\bibitem[{\citenamefont{Zhang et~al.}(2012)\citenamefont{Zhang, Li, Qi, Hu,
  Peng, Huang, and Wang}}]{ZLQ12}
\bibinfo{author}{\bibfnamefont{R.~L.} \bibnamefont{Zhang}},
  \bibinfo{author}{\bibfnamefont{J.}~\bibnamefont{Li}},
  \bibinfo{author}{\bibfnamefont{D.~X.} \bibnamefont{Qi}},
  \bibinfo{author}{\bibfnamefont{Q.}~\bibnamefont{Hu}},
  \bibinfo{author}{\bibfnamefont{R.~W.} \bibnamefont{Peng}},
  \bibinfo{author}{\bibfnamefont{R.~S.} \bibnamefont{Huang}}, \bibnamefont{and}
  \bibinfo{author}{\bibfnamefont{M.}~\bibnamefont{Wang}}, \bibinfo{journal}{J.
  Appl. Phys.} \textbf{\bibinfo{volume}{111}}, \bibinfo{pages}{07C325}
  (\bibinfo{year}{2012}).

\bibitem[{\citenamefont{Li et~al.}(2013)\citenamefont{Li, Ren, Zheng, Zhou,
  Wan, and Hao}}]{LRZZ13}
\bibinfo{author}{\bibfnamefont{S.}~\bibnamefont{Li}},
  \bibinfo{author}{\bibfnamefont{Z.}~\bibnamefont{Ren}},
  \bibinfo{author}{\bibfnamefont{J.}~\bibnamefont{Zheng}},
  \bibinfo{author}{\bibfnamefont{Y.}~\bibnamefont{Zhou}},
  \bibinfo{author}{\bibfnamefont{Y.}~\bibnamefont{Wan}}, \bibnamefont{and}
  \bibinfo{author}{\bibfnamefont{L.}~\bibnamefont{Hao}}, \bibinfo{journal}{J.
  Appl. Phys.} \textbf{\bibinfo{volume}{113}}, \bibinfo{pages}{033703}
  (\bibinfo{year}{2013}).

\bibitem[{\citenamefont{Koga et~al.}(2002)\citenamefont{Koga, Nitta, Akazaki,
  and Takayanagi}}]{KNAT02}
\bibinfo{author}{\bibfnamefont{T.}~\bibnamefont{Koga}},
  \bibinfo{author}{\bibfnamefont{J.}~\bibnamefont{Nitta}},
  \bibinfo{author}{\bibfnamefont{T.}~\bibnamefont{Akazaki}}, \bibnamefont{and}
  \bibinfo{author}{\bibfnamefont{H.}~\bibnamefont{Takayanagi}},
  \bibinfo{journal}{Phys. Rev. Lett.} \textbf{\bibinfo{volume}{89}},
  \bibinfo{pages}{046801} (\bibinfo{year}{2002}).

\bibitem[{\citenamefont{Bergsten et~al.}(2006)\citenamefont{Bergsten,
  Kobayashi, Sekine, and Nitta}}]{BKSN06}
\bibinfo{author}{\bibfnamefont{T.}~\bibnamefont{Bergsten}},
  \bibinfo{author}{\bibfnamefont{T.}~\bibnamefont{Kobayashi}},
  \bibinfo{author}{\bibfnamefont{Y.}~\bibnamefont{Sekine}}, \bibnamefont{and}
  \bibinfo{author}{\bibfnamefont{J.}~\bibnamefont{Nitta}},
  \bibinfo{journal}{Phys. Rev. Lett.} \textbf{\bibinfo{volume}{97}},
  \bibinfo{pages}{196803} (\bibinfo{year}{2006}).

\bibitem[{\citenamefont{Nagasawa et~al.}(2012)\citenamefont{Nagasawa, Takagi,
  Kunihashi, Kohda, and Nitta}}]{NTKKN12}
\bibinfo{author}{\bibfnamefont{F.}~\bibnamefont{Nagasawa}},
  \bibinfo{author}{\bibfnamefont{J.}~\bibnamefont{Takagi}},
  \bibinfo{author}{\bibfnamefont{Y.}~\bibnamefont{Kunihashi}},
  \bibinfo{author}{\bibfnamefont{M.}~\bibnamefont{Kohda}}, \bibnamefont{and}
  \bibinfo{author}{\bibfnamefont{J.}~\bibnamefont{Nitta}},
  \bibinfo{journal}{Phys. Rev. Lett.} \textbf{\bibinfo{volume}{108}},
  \bibinfo{pages}{086801} (\bibinfo{year}{2012}).

\bibitem[{\citenamefont{Nagasawa et~al.}(2013)\citenamefont{Nagasawa,
  Frustaglia, Saarikoski, Richter, and Nitta}}]{NFSRN13}
\bibinfo{author}{\bibfnamefont{F.}~\bibnamefont{Nagasawa}},
  \bibinfo{author}{\bibfnamefont{D.}~\bibnamefont{Frustaglia}},
  \bibinfo{author}{\bibfnamefont{H.}~\bibnamefont{Saarikoski}},
  \bibinfo{author}{\bibfnamefont{K.}~\bibnamefont{Richter}}, \bibnamefont{and}
  \bibinfo{author}{\bibfnamefont{J.}~\bibnamefont{Nitta}},
  \bibinfo{journal}{Nature Comm.} \textbf{\bibinfo{volume}{4}}
  (\bibinfo{year}{2013}).

\bibitem[{\citenamefont{F\"{o}ldi et~al.}(2010)\citenamefont{F\"{o}ldi,
  Szaszk\'{o}-Bog\'{a}r, and Peeters}}]{FSP10}
\bibinfo{author}{\bibfnamefont{P.}~\bibnamefont{F\"{o}ldi}},
  \bibinfo{author}{\bibfnamefont{V.}~\bibnamefont{Szaszk\'{o}-Bog\'{a}r}},
  \bibnamefont{and} \bibinfo{author}{\bibfnamefont{F.~M.}
  \bibnamefont{Peeters}}, \bibinfo{journal}{Phys. Rev. B}
  \textbf{\bibinfo{volume}{82}} (\bibinfo{year}{2010}).

\bibitem[{\citenamefont{F\"oldi et~al.}(2011)\citenamefont{F\"oldi,
  Szaszk\'o-Bog\'ar, and Peeters}}]{FSP11}
\bibinfo{author}{\bibfnamefont{P.}~\bibnamefont{F\"oldi}},
  \bibinfo{author}{\bibfnamefont{V.}~\bibnamefont{Szaszk\'o-Bog\'ar}},
  \bibnamefont{and} \bibinfo{author}{\bibfnamefont{F.~M.}
  \bibnamefont{Peeters}}, \bibinfo{journal}{Phys. Rev. B}
  \textbf{\bibinfo{volume}{83}}, \bibinfo{pages}{115313}
  (\bibinfo{year}{2011}).

\bibitem[{\citenamefont{Wu and Cao}(2007)}]{WC07}
\bibinfo{author}{\bibfnamefont{B.~H.} \bibnamefont{Wu}} \bibnamefont{and}
  \bibinfo{author}{\bibfnamefont{J.~C.} \bibnamefont{Cao}},
  \bibinfo{journal}{Phys. Rev. B} \textbf{\bibinfo{volume}{75}},
  \bibinfo{pages}{113303} (\bibinfo{year}{2007}).

\bibitem[{\citenamefont{F\"oldi et~al.}(2009)\citenamefont{F\"oldi, Benedict,
  K\'alm\'an, and Peeters}}]{FBKP09}
\bibinfo{author}{\bibfnamefont{P.}~\bibnamefont{F\"oldi}},
  \bibinfo{author}{\bibfnamefont{M.~G.} \bibnamefont{Benedict}},
  \bibinfo{author}{\bibfnamefont{O.}~\bibnamefont{K\'alm\'an}},
  \bibnamefont{and} \bibinfo{author}{\bibfnamefont{F.~M.}
  \bibnamefont{Peeters}}, \bibinfo{journal}{Phys. Rev. B}
  \textbf{\bibinfo{volume}{80}}, \bibinfo{pages}{165303}
  (\bibinfo{year}{2009}).

\bibitem[{\citenamefont{Romeo et~al.}(2008)\citenamefont{Romeo, Citro, and
  Marinaro}}]{RCM08}
\bibinfo{author}{\bibfnamefont{F.}~\bibnamefont{Romeo}},
  \bibinfo{author}{\bibfnamefont{R.}~\bibnamefont{Citro}}, \bibnamefont{and}
  \bibinfo{author}{\bibfnamefont{M.}~\bibnamefont{Marinaro}},
  \bibinfo{journal}{Phys. Rev. B} \textbf{\bibinfo{volume}{78}},
  \bibinfo{pages}{245309} (\bibinfo{year}{2008}).

\bibitem[{\citenamefont{Floquet}(1883)}]{F883}
\bibinfo{author}{\bibfnamefont{G.}~\bibnamefont{Floquet}},
  \bibinfo{journal}{Ann. \'{E}cole Norm. Sup.} \textbf{\bibinfo{volume}{12}},
  \bibinfo{pages}{46} (\bibinfo{year}{1883}).

\bibitem[{\citenamefont{Li and Reichl}(1999)}]{LR99}
\bibinfo{author}{\bibfnamefont{W.}~\bibnamefont{Li}} \bibnamefont{and}
  \bibinfo{author}{\bibfnamefont{L.~E.} \bibnamefont{Reichl}},
  \bibinfo{journal}{Phys. Rev. B} \textbf{\bibinfo{volume}{60}},
  \bibinfo{pages}{15732} (\bibinfo{year}{1999}).

\bibitem[{\citenamefont{L\'opez et~al.}(2012)\citenamefont{L\'opez, Sun, and
  Schliemann}}]{LSS12}
\bibinfo{author}{\bibfnamefont{A.}~\bibnamefont{L\'opez}},
  \bibinfo{author}{\bibfnamefont{Z.~Z.} \bibnamefont{Sun}}, \bibnamefont{and}
  \bibinfo{author}{\bibfnamefont{J.}~\bibnamefont{Schliemann}},
  \bibinfo{journal}{Phys. Rev. B} \textbf{\bibinfo{volume}{85}},
  \bibinfo{pages}{205428} (\bibinfo{year}{2012}).

\bibitem[{\citenamefont{Moskalets}(2014)}]{M14}
\bibinfo{author}{\bibfnamefont{M.}~\bibnamefont{Moskalets}},
  \bibinfo{journal}{Phys. Rev. Lett.} \textbf{\bibinfo{volume}{112}},
  \bibinfo{pages}{206801} (\bibinfo{year}{2014}).

\bibitem[{\citenamefont{Dasenbrook et~al.}(2014)\citenamefont{Dasenbrook,
  Flindt, and B\"uttiker}}]{DFB14}
\bibinfo{author}{\bibfnamefont{D.}~\bibnamefont{Dasenbrook}},
  \bibinfo{author}{\bibfnamefont{C.}~\bibnamefont{Flindt}}, \bibnamefont{and}
  \bibinfo{author}{\bibfnamefont{M.}~\bibnamefont{B\"uttiker}},
  \bibinfo{journal}{Phys. Rev. Lett.} \textbf{\bibinfo{volume}{112}},
  \bibinfo{pages}{146801} (\bibinfo{year}{2014}).

\bibitem[{\citenamefont{Nowak et~al.}(2012)\citenamefont{Nowak, Szafran, and
  Peeters}}]{NSzP012}
\bibinfo{author}{\bibfnamefont{M.~P.} \bibnamefont{Nowak}},
  \bibinfo{author}{\bibfnamefont{B.}~\bibnamefont{Szafran}}, \bibnamefont{and}
  \bibinfo{author}{\bibfnamefont{F.~M.} \bibnamefont{Peeters}},
  \bibinfo{journal}{Phys. Rev. B} \textbf{\bibinfo{volume}{86}},
  \bibinfo{pages}{125428} (\bibinfo{year}{2012}).

\bibitem[{\citenamefont{Szaszk\'o-Bog\'ar
  et~al.}(2014)\citenamefont{Szaszk\'o-Bog\'ar, F\"oldi, and Peeters}}]{SFP14}
\bibinfo{author}{\bibfnamefont{V.}~\bibnamefont{Szaszk\'o-Bog\'ar}},
  \bibinfo{author}{\bibfnamefont{P.}~\bibnamefont{F\"oldi}}, \bibnamefont{and}
  \bibinfo{author}{\bibfnamefont{F.~M.} \bibnamefont{Peeters}},
  \bibinfo{journal}{J. Phys: Cond. Mat.} \textbf{\bibinfo{volume}{26}},
  \bibinfo{pages}{135302} (\bibinfo{year}{2014}).

\bibitem[{\citenamefont{Sadreev and Sherman}(2013)}]{SS13}
\bibinfo{author}{\bibfnamefont{A.~F.} \bibnamefont{Sadreev}} \bibnamefont{and}
  \bibinfo{author}{\bibfnamefont{E.~Y.} \bibnamefont{Sherman}},
  \bibinfo{journal}{Phys. Rev. B} \textbf{\bibinfo{volume}{88}},
  \bibinfo{pages}{115302} (\bibinfo{year}{2013}).

\bibitem[{\citenamefont{Paw³owski et~al.}(2014)\citenamefont{Paw³owski,
  Szumniak, Skubis, and Bednarek}}]{PSzSB14}
\bibinfo{author}{\bibfnamefont{J.}~\bibnamefont{Paw³owski}},
  \bibinfo{author}{\bibfnamefont{P.}~\bibnamefont{Szumniak}},
  \bibinfo{author}{\bibfnamefont{A.}~\bibnamefont{Skubis}}, \bibnamefont{and}
  \bibinfo{author}{\bibfnamefont{S.}~\bibnamefont{Bednarek}},
  \bibinfo{journal}{J. Phys: Cond. Mat.} \textbf{\bibinfo{volume}{26}},
  \bibinfo{pages}{345302} (\bibinfo{year}{2014}).

\bibitem[{\citenamefont{F\"oldi et~al.}(2010)\citenamefont{F\"oldi, K\'alm\'an,
  and Benedict}}]{FKB10}
\bibinfo{author}{\bibfnamefont{P.}~\bibnamefont{F\"oldi}},
  \bibinfo{author}{\bibfnamefont{O.}~\bibnamefont{K\'alm\'an}},
  \bibnamefont{and} \bibinfo{author}{\bibfnamefont{M.~G.}
  \bibnamefont{Benedict}}, \bibinfo{journal}{Phys. Rev. B}
  \textbf{\bibinfo{volume}{82}}, \bibinfo{pages}{165322}
  (\bibinfo{year}{2010}).

\bibitem[{\citenamefont{Thorgilsson et~al.}(2012)\citenamefont{Thorgilsson,
  Egues, Loss, and Erlingsson}}]{TELE12}
\bibinfo{author}{\bibfnamefont{G.}~\bibnamefont{Thorgilsson}},
  \bibinfo{author}{\bibfnamefont{J.~C.} \bibnamefont{Egues}},
  \bibinfo{author}{\bibfnamefont{D.}~\bibnamefont{Loss}}, \bibnamefont{and}
  \bibinfo{author}{\bibfnamefont{S.~I.} \bibnamefont{Erlingsson}},
  \bibinfo{journal}{Phys. Rev. B} \textbf{\bibinfo{volume}{85}},
  \bibinfo{pages}{045306} (\bibinfo{year}{2012}).

\bibitem[{\citenamefont{Meijer et~al.}(2002)\citenamefont{Meijer, Morpurgo, and
  Klapwijk}}]{MMK02}
\bibinfo{author}{\bibfnamefont{F.~E.} \bibnamefont{Meijer}},
  \bibinfo{author}{\bibfnamefont{A.~F.} \bibnamefont{Morpurgo}},
  \bibnamefont{and} \bibinfo{author}{\bibfnamefont{T.~M.}
  \bibnamefont{Klapwijk}}, \bibinfo{journal}{Phys. Rev. B}
  \textbf{\bibinfo{volume}{66}}, \bibinfo{pages}{033107}
  (\bibinfo{year}{2002}).

\bibitem[{\citenamefont{Krstaji\'{c} et~al.}(2011)\citenamefont{Krstaji\'{c},
  Pagano, and Vasilopoulos}}]{KPV11}
\bibinfo{author}{\bibfnamefont{P.}~\bibnamefont{Krstaji\'{c}}},
  \bibinfo{author}{\bibfnamefont{M.}~\bibnamefont{Pagano}}, \bibnamefont{and}
  \bibinfo{author}{\bibfnamefont{P.}~\bibnamefont{Vasilopoulos}},
  \bibinfo{journal}{Phys. E} \textbf{\bibinfo{volume}{43}},
  \bibinfo{pages}{893} (\bibinfo{year}{2011}).

\bibitem[{\citenamefont{Zhang and Zhu}(2013)}]{ZZ13}
\bibinfo{author}{\bibfnamefont{S.-f.} \bibnamefont{Zhang}} \bibnamefont{and}
  \bibinfo{author}{\bibfnamefont{W.}~\bibnamefont{Zhu}}, \bibinfo{journal}{J.
  Phys. : Condens. Matter} \textbf{\bibinfo{volume}{25}},
  \bibinfo{pages}{075302} (\bibinfo{year}{2013}).

\bibitem[{\citenamefont{Szafran}(2011)}]{Sz11}
\bibinfo{author}{\bibfnamefont{B.}~\bibnamefont{Szafran}},
  \bibinfo{journal}{Phys. Rev. B} \textbf{\bibinfo{volume}{84}},
  \bibinfo{pages}{075336} (\bibinfo{year}{2011}).

\bibitem[{\citenamefont{Pan and Zhao}(2012)}]{PZ12}
\bibinfo{author}{\bibfnamefont{H.}~\bibnamefont{Pan}} \bibnamefont{and}
  \bibinfo{author}{\bibfnamefont{Y.}~\bibnamefont{Zhao}}, \bibinfo{journal}{J.
  Appl. Phys.} \textbf{\bibinfo{volume}{111}}, \bibinfo{pages}{083703}
  (\bibinfo{year}{2012}).

\bibitem[{\citenamefont{Shakouri et~al.}(2012)\citenamefont{Shakouri, Szafran,
  Esmaeilzadeh, and Peeters}}]{SSzEP12}
\bibinfo{author}{\bibfnamefont{K.}~\bibnamefont{Shakouri}},
  \bibinfo{author}{\bibfnamefont{B.}~\bibnamefont{Szafran}},
  \bibinfo{author}{\bibfnamefont{M.}~\bibnamefont{Esmaeilzadeh}},
  \bibnamefont{and} \bibinfo{author}{\bibfnamefont{F.~M.}
  \bibnamefont{Peeters}}, \bibinfo{journal}{Phys. Rev. B}
  \textbf{\bibinfo{volume}{85}}, \bibinfo{pages}{165314}
  (\bibinfo{year}{2012}).

\bibitem[{\citenamefont{Maiti}(2015)}]{Ma15}
\bibinfo{author}{\bibfnamefont{S.~K.} \bibnamefont{Maiti}},
  \bibinfo{journal}{J. Appl. Phys.} \textbf{\bibinfo{volume}{117}}
  (\bibinfo{year}{2015}).

\bibitem[{\citenamefont{K\'{a}lm\'{a}n
  et~al.}(2008)\citenamefont{K\'{a}lm\'{a}n, F\"{o}ldi, Benedict, and
  Peeters}}]{KFBP08b}
\bibinfo{author}{\bibfnamefont{O.}~\bibnamefont{K\'{a}lm\'{a}n}},
  \bibinfo{author}{\bibfnamefont{P.}~\bibnamefont{F\"{o}ldi}},
  \bibinfo{author}{\bibfnamefont{M.~G.} \bibnamefont{Benedict}},
  \bibnamefont{and} \bibinfo{author}{\bibfnamefont{F.~M.}
  \bibnamefont{Peeters}}, \bibinfo{journal}{Phys. Rev. B}
  \textbf{\bibinfo{volume}{78}}, \bibinfo{pages}{125306}
  (\bibinfo{year}{2008}).

\bibitem[{\citenamefont{F\"{o}ldi et~al.}(2008)\citenamefont{F\"{o}ldi,
  K\'{a}lm\'{a}n, Benedict, and Peeters}}]{KFBP08c}
\bibinfo{author}{\bibfnamefont{P.}~\bibnamefont{F\"{o}ldi}},
  \bibinfo{author}{\bibfnamefont{O.}~\bibnamefont{K\'{a}lm\'{a}n}},
  \bibinfo{author}{\bibfnamefont{M.~G.} \bibnamefont{Benedict}},
  \bibnamefont{and} \bibinfo{author}{\bibfnamefont{F.~M.}
  \bibnamefont{Peeters}}, \bibinfo{journal}{Nano.~Lett.}
  \textbf{\bibinfo{volume}{8}}, \bibinfo{pages}{2556} (\bibinfo{year}{2008}).

\bibitem[{\citenamefont{Eslami and Esmaeilzadeh}(2014)}]{EE14}
\bibinfo{author}{\bibfnamefont{L.}~\bibnamefont{Eslami}} \bibnamefont{and}
  \bibinfo{author}{\bibfnamefont{M.}~\bibnamefont{Esmaeilzadeh}},
  \bibinfo{journal}{J. Appl. Phys.} \textbf{\bibinfo{volume}{115}},
  \bibinfo{pages}{084307} (\bibinfo{year}{2014}).

\bibitem[{\citenamefont{Abramowitz and Stegun}(1965)}]{AS65}
\bibinfo{editor}{\bibfnamefont{M.}~\bibnamefont{Abramowitz}} \bibnamefont{and}
  \bibinfo{editor}{\bibfnamefont{I.}~\bibnamefont{Stegun}}, eds.,
  \emph{\bibinfo{title}{Handbook of mathematical functions}}
  (\bibinfo{publisher}{Dover Publications}, \bibinfo{address}{New York},
  \bibinfo{year}{1965}).

\bibitem[{\citenamefont{Griffith}(1953)}]{G53}
\bibinfo{author}{\bibfnamefont{S.}~\bibnamefont{Griffith}},
  \bibinfo{journal}{Trans. Faraday Soc.} \textbf{\bibinfo{volume}{49}},
  \bibinfo{pages}{345} (\bibinfo{year}{1953}).

\bibitem[{\citenamefont{{P. F\"{o}ldi} et~al.}(2001)\citenamefont{{P.
  F\"{o}ldi}, Czirj\'{a}k, and Benedict}}]{FCB01a}
\bibinfo{author}{\bibnamefont{{P. F\"{o}ldi}}},
  \bibinfo{author}{\bibfnamefont{A.}~\bibnamefont{Czirj\'{a}k}},
  \bibnamefont{and} \bibinfo{author}{\bibfnamefont{M.~G.}
  \bibnamefont{Benedict}}, \bibinfo{journal}{Phys. Rev. A}
  \textbf{\bibinfo{volume}{63}}, \bibinfo{pages}{033807}
  (\bibinfo{year}{2001}).

\bibitem[{\citenamefont{Kiselev and Kim}(2005)}]{KK05}
\bibinfo{author}{\bibfnamefont{A.~A.} \bibnamefont{Kiselev}} \bibnamefont{and}
  \bibinfo{author}{\bibfnamefont{K.~W.} \bibnamefont{Kim}},
  \bibinfo{journal}{Phys. Rev. B} \textbf{\bibinfo{volume}{71}},
  \bibinfo{pages}{153315} (\bibinfo{year}{2005}).

\bibitem[{\citenamefont{Moln\'{a}r et~al.}(2004)\citenamefont{Moln\'{a}r,
  Peeters, and Vasilopoulos}}]{MPV04}
\bibinfo{author}{\bibfnamefont{B.}~\bibnamefont{Moln\'{a}r}},
  \bibinfo{author}{\bibfnamefont{F.~M.} \bibnamefont{Peeters}},
  \bibnamefont{and}
  \bibinfo{author}{\bibfnamefont{P.}~\bibnamefont{Vasilopoulos}},
  \bibinfo{journal}{Phys. Rev. B} \textbf{\bibinfo{volume}{69}},
  \bibinfo{pages}{155335} (\bibinfo{year}{2004}).

\end{thebibliography}
\end{document}